\documentclass[12pt]{article}
\usepackage{amssymb}
\usepackage{amsmath}
\usepackage{graphicx}

\newtheorem{theor}{Theorem}
\newcommand{\be}{\begin{equation}}
\newcommand{\ee}{\end{equation}}
\newcommand{\bea}{\begin{eqnarray}}
\newcommand{\eea}{\end{eqnarray}}
\DeclareMathOperator{\sech}{sech}
\begin{document}

\title{\textbf{Null Energy Condition Violation and Classical Stability in the Bianchi I Metric}}

\author{I.~Ya.~Aref'eva$^1$\footnote{arefeva@mi.ras.ru},
N.~V.~Bulatov$^2$\footnote{nick\_bulatov@mail.ru},
L.~V.~Joukovskaya$^3$\footnote{l.joukovskaya@damtp.cam.ac.uk},
S.~Yu.~Vernov$^4$\footnote{svernov@theory.sinp.msu.ru} \\[2.7mm]
${}^1$\small{Steklov Mathematical Institute, Russian Academy of Sciences,}\\
\small{Gubkina str. 8, 119991, Moscow, Russia}\\
${}^2$\small{Department of Quantum Statistics and Field Theory, Faculty
of Physics,}\\ \small{Moscow State University, Leninskie
Gory 1, 119991, Moscow, Russia,}\\
${}^3$\small{Centre for Theoretical Cosmology, DAMTP, CMS, University
of Cambridge},\\ \small{ Wilberforce Road, CB3 0WA, Cambridge, United
Kingdom}\\
${}^4$\small{Skobeltsyn Institute of Nuclear Physics, Moscow State University},\\
\small{Leninskie Gory 1, 119991, Moscow, Russia}\\
}

\date{ }

\maketitle

\begin{abstract}
The stability of isotropic cosmological solutions in the Bianchi I
model is considered. We prove that the stability of isotropic solutions
in the Bianchi I metric for a positive Hubble parameter follows from
their stability in the Friedmann--Robertson--Walker metric. This result
is applied to models inspired by string field theory, which violate the
null energy condition. Examples of stable isotropic solutions are
presented. We also consider the $k$-essence model and analyse the
stability of solutions of the form $\Phi(t)=t$.
\end{abstract}

\section{Introduction}
Field theories which violate the null energy condition (NEC) are of
interest for the solution of the cosmological singularity problem
\cite{Hawking-Ellis,cyclic,GV} and for models of dark energy with the
equation of state parameter $w<-1$ (see \cite{Caldwell}--\cite{Carroll}
and references therein).  Generally speaking, models that violate the
NEC have ghosts, and therefore  are unstable and physically
unacceptable.

However, the possibility of the existence of dark energy with $w<-1$
 on the one hand\footnote{This possibility is  not excluded experimentally
\cite{cosmo-obser}, see \cite{DE1,Albrecht} for reviews of dynamical
dark energy models.} and the cosmological singularity problem on the
other hand encourage the investigation of models which violate the NEC.
It is almost clear that such a possibility can be realized within an
effective theory, while the fundamental  theory should  be stable and
admit quantization. From this point of view the NEC violation might be
a property of a model that approximates the fundamental theory and
describes some particular features of the fundamental theory. With the
lack of quantum gravity, we can just trust string theory or deal with
an effective theory admitting  the UV completion.

There have been several attempts  to realize these scenarios
\cite{GostCond,IA,Creminelli}. The ghost condensation model
\cite{GostCond,Mukohyama,EFTinflation,Creminelli-0702165} proposed to
describe a wide class of cosmological perturbations has a ghost in the
perturbative vacuum and has no ghost in the ghost condensation phase
within an effective theory. The new ekpyrotic scenario
\cite{Creminelli,Buchbinder-0702154,Buchbinder-0706.3903,Creminelli-eff}
is a development of the ekpyrotic \cite{KOST} and the  cyclic scenarios
\cite{cyclic,Craps}, and it attempts to solve the singularity problem,
among others, by involving violation of the NEC. Nonlocal cosmological
models \cite{IA,AJK,AJ,Calcagni,Barnaby,AK,AJV1,LJ-PR,AJV2,Mulruny,KV}
inspired by the string field theory (SFT) \cite{Witten,NPB,SFT-review}
admit a regime with $w<-1$.

All of these models possess higher derivatives terms, which produce
well-known problems with quantum instability \cite{AV-NEC,RAS}. Several
attempts to solve these problems have been recently performed
\cite{SW,Creminelli-eff}.  A  physical idea that could
 solve the problems
is  that  the instabilities do not have  enough time to fully develop.
A mathematical one is that dangerous terms can be treated as
corrections valued only at small energies below the physical cut-off.
This approach implies the possibility  to construct a UV completion of
the theory, and this assumption requires detailed analysis.

The NEC plays an important role in  classical general relativity, in
particular, in the consideration of  black holes and cosmological
singularities \cite{Hawking-Ellis,GV}.

The NEC violating models can admit classically  stable solutions in the
Friedmann--Robertson--Walker (FRW) cosmology. In particular,  there are
classically stable solutions for self-interacting  ghost models with
minimal coupling to  gravity. Moreover,  there exists an attractor
behavior (for details about attractor solutions for inhomogeneous
cosmological models, see~\cite{Starobinsky}) in a class of the phantom
cosmological models \cite{AKVCDM,phantom-attractor,Lazkoz}. One can
study the stability of the FRW metric, specifying a form of
fluctuations. It is interesting to know whether these solutions are
stable under the deformation of the FRW metric to an anisotropic one,
for example, to the Bianchi I metric. In comparison with  general
fluctuations we can get an explicit form of solutions in the Bianchi I
metric, which can probably clarify some nontrivial issues of  theories
with NEC violation.

Stability of isotropic solutions in the Bianchi
models~\cite{Bianchi,Ellis98,Ellis} (see also~\cite{DSC}) has been
considered in inflationary models (see~\cite{Germani,Koivisto} and
references therein for details of anisotropic slow-roll inflation).
Assuming that the energy conditions are satisfied, it has been proved
that all initially expanding Bianchi models except type IX approach the
de Sitter space-time~\cite{Wald} (see
also~\cite{MossSahni,Wainwright,Kitada,Rendall04}).   The Wald
theorem~\cite{Wald} shows that for space-time of Bianchi types I--VIII
with a positive cosmological constant and matter satisfying the
dominant and strong energy conditions, solutions which exist globally
in the future have certain asymptotic properties at
$t\rightarrow\infty$. It is interesting to consider a similar question
in the case of phantom cosmology \cite{phantom-attractor,AKV,Lazkoz}
and string inspired models \cite{IA,AJ,AJV1,AJV2,AKV2,Vernov06}, as
well as in the case of the ghost condensation models \cite{GostCond} or
their modifications \cite{Creminelli-eff}.

The Bianchi universe models \cite{Bianchi,Ellis98,Ellis} are spatially
homogeneous anisotropic cosmological models. There are strong limits on
anisotropic models from observations \cite{Barrow,Bernui:2005pz}.
Anisotropic spatially homogeneous fluctuations have to be strongly
suppressed, and models developing large anisotropy should be discarded
as early or late cosmological models.

In this paper we consider the stability of isotropic solutions in the
Bianchi I metric in the presence of phantom scalar fields. There are
two classes of models whose stability we analyse in this paper. The
first class includes the  one phantom scalar field models of  dark
energy, which admit exact kink-type or lump-type
solutions~\cite{AKV,AJ}. For this class of models we also analyse the
stability with respect to small fluctuations of the initial value of
the cold dark matter energy density (compare with \cite{AKVCDM}). The
second class includes models with a scalar field $\phi$, which have
exact solutions $\phi\sim t$, for example, the $k$-essence
models~\cite{Mukhanov,Malquarti,Chimento,Sen}, in particular, ghost
condensate models~\cite{GostCond,Mukohyama,Creminelli-eff}.

For both classes of models we prove that the stability of the solutions
in the Bianchi I metric is  equivalent to the stability of the
corresponding solutions in the FRW metric. The stability of a kink or
lump solution in the FRW metric means the stability of the fixed point
that the solution tends to. Using the Lyapunov theorem
\cite{Lyapunov,Pontryagin} we find conditions under which the fixed
point and the corresponding kink (or lump) solution are stable. In
these cases the necessary condition for the exact solution's stability
is boundedness of the first corrections for the positive time semiaxis.
When we can not use this theorem  we check the boundedness of the first
corrections to the exact solutions explicitly.

The paper is organized as follows. In Section 2 we deal with an
arbitrary $N$-component scalar potential model and a $k$-essence model
in the Bianchi I metric. We also review the Lyapunov theorem and other
important statement about stability. In Section 3 we consider the
stability of solutions which tend to an isolated fixed point in
one-field models with the cold dark matter (CDM). We find  sufficient
conditions for the stability of such solutions in the FRW and Bianchi I
metrics. In Section 4 we find  the connection between the first order
corrections in the FRW and Bianchi I metrics for $N$-field models. The
corresponding result for the $k$-essence model is presented in Section
5. In Section 6 we present examples of stable isotropic kink and lump
solutions in SFT inspired cosmological models. We also analyse the
first order corrections for solutions, which are proportional to time.
In Section 7 we make a conclusion and propose directions for further
investigations.

\section{Setup}

\subsection{The Bianchi I cosmological model with scalar and phantom scalar fields and the CDM}

Let us start with a cosmological model with $N$ scalar fields $\phi_1$,
$\phi_2$, $\dots$, $\phi_N$ in the  Bianchi I metric
\begin{equation}
\label{Bianchi}
{ds}^{2}={}-{dt}^2+a_1^2(t)dx_1^2+a_2^2(t)dx_2^2+a_3^2(t)dx_3^2.
\end{equation}
The action is
\begin{equation}
 S=\int d^4x \sqrt{-g}\left(\frac{R}{16\pi G_N}-
 \sum _{k=1}^{N}\frac{C_k}{2}g^{\mu\nu}\partial_{\mu}\phi_k\partial_{\nu}\phi_k
-V(\phi_1,\dots,\phi_N)-\Lambda\right), \label{action_N}
\end{equation}
where the potential $V$ is a twice continuously differentiable
function, $G_N$ is the Newtonian gravitational constant, $\Lambda$ is a
cosmological constant, and $C_k$ are nonzero real numbers. The sign of
$C_k$ defines whether field $\phi_k$ is the phantom field ($C_k<0$) or
the ordinary scalar field ($C_k>0$).

The Einstein equations have the following form:
\begin{equation}
\label{a} H_1H_2+H_1H_3+H_2H_3=8\pi G_N\varrho,
\end{equation}
\begin{equation}
\label{b} \dot H_2+H_2^2+\dot H_3+H_3^2+H_2H_3={}-8\pi G_N p,
\end{equation}
\begin{equation}
\label{c} \dot H_1+H_1^2+\dot H_2+H_2^2+H_1H_2={}-8\pi G_N p,
\end{equation}
\begin{equation}
\label{d} \dot H_1+H_1^2+\dot H_3+H_3^2+H_1H_3={}-8\pi G_N p,
\end{equation}
where
\begin{equation}
\label{varrho} \varrho=\sum\limits_{k=1}^N
    \frac{C_k}{2}\dot{\phi}_k^2+V(\phi_1,\dots,\phi_N)+\Lambda+\rho_m,
\end{equation}
\begin{equation}
\label{pressure}
    p=\sum\limits_{k=1}^N
    \frac{C_k}{2}\dot{\phi}_k^2-V(\phi_1,\dots,\phi_N)-\Lambda,
\end{equation}
\begin{equation}
H_1=\frac{\dot a_1}{a_1},\qquad H_2=\frac{\dot a_2}{a_2}, \qquad
H_3=\frac{\dot a_3}{a_3}
\end{equation}
and a dot denotes a time derivative.

Note that we couple, in a minimal way, pressureless matter (the CDM)
with the energy density $\rho_{m}$  to our model. The equation for the
CDM energy density is as follows:
\begin{equation}
\label{rhoequ} \dot\rho_m={}-(H_1+H_2+H_3)\rho_m.
\end{equation}

Introducing $\psi_k=\dot\phi_{k}$ we obtain from action
(\ref{action_N}) the following equations:
\begin{equation}
\label{e}
\begin{array}{l}
\displaystyle \dot \phi_k =\psi_{k},\\
\displaystyle
\dot{\psi}_k={}-(H_1+H_2+H_3)\psi_k-\frac{1}{C_k}V'_{\phi_{k}},
\end{array}
\end{equation}
where $V'_{\phi_{k}}\equiv \frac{\partial V}{\partial\phi_k}$,
$k=1,2,\dots,N$. Thus we get the system of $2N+4$ first order
differential equations and  one constraint~(\ref{a}).

It is convenient to express the initial variables $a_i$ in terms of new
variables $a$ and $\beta_i$ (we use notations from~\cite{Pereira}),
subject to the following constraint:
\begin{equation}
\label{restr1} \beta_1+\beta_2+\beta_3=0.
\end{equation}

One has the following relations
\begin{equation}
a_i(t)= a(t) e^{\beta_i(t)}, \quad\mbox{hence,}\quad
a(t)=(a_1(t)a_2(t)a_3(t))^{1/3},
\end{equation}
\begin{equation}\label{Hi}
H_i\equiv H+\dot\beta_i, \qquad\mbox{and}\qquad
H=\frac{1}{3}(H_1+H_2+H_3),
\end{equation}
where $H\equiv \dot a/a$. To obtain (\ref{Hi}) we have used the
following consequence of (\ref{restr1}):
\begin{equation}
\label{restr2} \dot\beta_1+\dot\beta_2+\dot\beta_3=0.
\end{equation}

Note that $\beta_i$ are not components of a vector and, therefore, are
not subjected to the Einstein summation rule. In the case of the FRW
metric all $\beta_i$ are equal to zero and $H$ is the Hubble parameter.
Following~\cite{Ellis98,Pereira} (see also~\cite{DSC}) we introduce the shear
\begin{equation}
\sigma^2\equiv \dot\beta_1^2+\dot\beta_2^2+\dot\beta_3^2.
\end{equation}
It is useful to write  equations (\ref{a})--(\ref{d}), (\ref{rhoequ})
and (\ref{e}) in terms of new variables.

Using relation (\ref{restr2}) we can write equation (\ref{a}) as
follows
\begin{equation}
\label{a2} 3H^2-\frac{1}{2}\sigma^2=8\pi G_N\varrho.
\end{equation}

Summing equations (\ref{b})--(\ref{d}) one can obtain
\begin{equation}
\label{trequ} 2\dot H+3H^2+\frac{1}{2}\sigma^2={}-8\pi G_Np.
\end{equation}

Therefore
\begin{equation}
\label{equaH2} \dot H+3H^2=4\pi G_N(\varrho-p).
\end{equation}

Note that equations (\ref{rhoequ}) and (\ref{e}) in new variables,
\begin{equation}
\label{e2}
\dot\phi_k=\psi_k,\qquad
\dot{\psi}_k={}-3H\psi_k-\frac{1}{C_k}V'_{\phi_{k}},
\end{equation}
\begin{equation}
\label{rhoequ2}
\dot\rho_m={}-3H\rho_m,
\end{equation}
 as well as equation (\ref{equaH2}), look like the corresponding equations in the
FRW metric.

Subtracting (\ref{b}) from (\ref{c}) we obtain
\begin{equation}
\label{c-b} \dot H_1+H_1^2-\dot H_3-H_3^2+H_2(H_1-H_3)=0.
\end{equation}
In terms of $H$ and $\beta_i$ equation (\ref{c-b}) takes the form
\begin{equation}
\ddot\beta_1+3H\dot \beta_1=\ddot\beta_3+3H\dot \beta_3. \label{c-b2}
\end{equation}

Using (\ref{c-b2}) and (\ref{restr2}) we obtain the following equations
\begin{equation}
\ddot\beta_i={}-3H\dot \beta_i, \label{equbeta}
\end{equation}
\begin{equation}
\label{equvartheta} \frac{d}{dt}\left(\sigma^2\right)={}-6H\sigma^2.
\end{equation}

Functions $H(t)$ and $\sigma^2(t)$ together with $\phi_k(t)$ and
$\rho_m(t)$ can be obtained from equations
(\ref{trequ})--(\ref{rhoequ2}) and (\ref{equvartheta}). If $H(t)$ is
known, then $\beta_i$ can be trivially obtained from (\ref{equbeta}).
We show in the next section that functions $H(t)$, $\dot\beta_i(t)$,
and $\sigma^2(t)$ are very suitable to analyse the stability of
isotropic solutions in the Bianchi I metric.

\subsection{$k$-essence model in the Bianchi I metric}
Let us consider the $k$-essence cosmological model, which is described
by the action
\begin{equation}
S=\int d^4x \sqrt{-g}\left(\frac{R}{16\pi G_N}-
 {\cal P}(\Phi,X)-\Lambda\right), \label{action_kEs}
\end{equation}
 where
\begin{equation} \label{X}
X\equiv{}-g^{\mu\nu}\partial_\mu\Phi\partial_\nu\Phi.
\end{equation}

The pressure ${\cal P}(\Phi,X)$ is of the
form~\cite{GostCond,Creminelli}
\begin{equation}
\label{Pke} {\cal
P}(\Phi,X)=\frac{1}{2}(p_q(\Phi)-\varrho_q(\Phi))+\frac{1}{2}(p_q(\Phi)+\varrho_q(\Phi))X
+\frac{1}{2}M^4(\Phi)(X-1)^2.
\end{equation}
Here $p_q(\Phi)$, $\varrho_q(\Phi)$, and $M^4(\Phi)$ are arbitrary
functions of $\Phi$. The energy density is
\begin{equation}
\label{Eke}
    {\cal E}(\Phi,X)=(p_q(\Phi)+\varrho_q(\Phi))X+2M^4(\Phi)(X^2-X)-{\cal P}(\Phi,X).
\end{equation}

In the  Bianchi I metric for $\Phi$, depending only on time, we have
$X=\dot\Phi^2$. The Einstein equations are
 \bea
 H_1H_2+H_1H_3+H_2H_3&=&8\pi G_N({\cal E}+\Lambda),\label{equKes1} \\
\dot H_2+H_2^2+\dot H_3+H_3^2+H_2H_3&=&{}-8\pi
G_N({\cal P}-\Lambda),\label{equKes2}\\
\dot H_1+H_1^2+\dot H_2+H_2^2+H_1H_2&=&{}-8\pi
G_N({\cal P}-\Lambda),\\
\dot H_1+H_1^3+\dot H_3+H_3^2+H_1H_3&=&{}-8\pi G_N({\cal P}-\Lambda).
\label{equKes4} \eea

From action (\ref{action_kEs}) we also obtain the second order
differential equation for the $k$-essence field $\Phi$, which
represents a consequence of system (\ref{equKes1})--(\ref{equKes4}).
Indeed, we differentiate (\ref{equKes1}) with respect to $t$ and obtain

\begin{equation}
(\dot H_2+\dot H_3)H_1+(\dot H_1+\dot H_3)H_2+(\dot H_1+\dot
H_2)H_3=8\pi G_N\dot{\cal E}.
\end{equation}
Using (\ref{equKes1})--(\ref{equKes4}) to exclude $\dot H_i$, we
transform this equation into the following form:
\begin{equation}
\dot {\cal E}={}-(H_1+H_2+H_3)\left({\cal E}+{\cal P}\right).
\end{equation}

Substituting explicit forms of ${\cal E}$ and ${\cal P}$, we obtain
\begin{equation}
\label{equPhi}
\begin{array}{lcl}
\displaystyle \left(2T_q+M^4(3\dot\Phi^2-1)\right)\ddot\Phi &=&
{}-T_q'\dot\Phi^2-V'_q
-2M^3M'\left(3\dot\Phi^4-2\dot\Phi^2-1\right) -{}\\
&\displaystyle -&\displaystyle
2(H_1+H_2+H_3)\dot\Phi\left(T_q+2M^4(\dot\Phi^2-1)\right),
\end{array}
\end{equation}
where a prime denotes a derivative with respect to $\Phi$,
\begin{equation}
V_q(\Phi)\equiv \frac{1}{2}(\varrho_q(\Phi)-p_q(\Phi)),\qquad
T_q(\Phi)\equiv \frac{1}{2}(p_q(\Phi)+\varrho_q(\Phi)).
\end{equation}

The $k$-essence model has one important property. For any real
differentiable function $H_0(t)$, there exist such  real differentiable
functions $\varrho_q(\Phi)$ and $p_q(\Phi)$ that the functions
$H_i(t)=H_0(t)$ and $\Phi(t)=t$ solve system
(\ref{equKes1})--(\ref{equKes4}) and, therefore, equation
(\ref{equPhi}). Indeed at $\Phi(t)=t$,
\begin{equation}
 {\cal E}=\varrho_q(\Phi)=\varrho_q(t),\qquad {\cal P}=p_q(\Phi)=p_q(t).
\end{equation}
So, one can obtain from (\ref{equKes1})--(\ref{equKes4})
\begin{equation}
\varrho_q(t)=\frac{3}{8\pi G_N}H_0^2(t)-\Lambda, \qquad
p_q(t)={}-\varrho_q(t)-\frac{1}{4\pi G_N}\dot H(t).
\end{equation}
Substituting the obtained $\varrho_q(\Phi)$ and $p_q(\Phi)$ in
(\ref{Pke}), we see that the system (\ref{equKes1})--(\ref{equKes4})
has a particular solution $H_i(t)=H_0(t)$ and $\Phi(t)=t$.

Bianchi--type models I--VIII coupled to  $k$-essence matter
representing dark energy and other matter which satisfies the strong
and dominant energy conditions have been considered
in~\cite{Rendall05}. A general criterion for isotropization of these
models has been derived~\cite{Rendall05}. In this paper we do not
assume that the energy conditions are satisfied when considering
$k$-essence models in the Bianchi I metric.

\subsection{A few known facts about stability}

Let us remember a few facts about the stability
\cite{Lyapunov,Pontryagin,Petrovsky} of solutions for a general system
of the first order autonomic  equations \begin{equation} \label{aue}
\dot y_k=F_k(y),\qquad k=1,2,\dots,N. \end{equation}

By definition a solution (a trajectory)
$y_0(t)$ is attractive (stable) if
\begin{equation}
\|\tilde{y}(t)-y_0(t)\| \rightarrow 0 \quad \mbox{at} \quad
t\rightarrow \infty
\end{equation}
for all solutions $\tilde{y}(t)$ that
start close enough to $y_0(t)$.

If all solutions of the dynamical system that start out near a fixed
(equilibrium) point $y_f$,
\begin{equation}
\label{auf} F_k(y_f)=0, \qquad k=1,2,\dots,N
\end{equation}
stay near $y_f$  forever, then $y_f$ is {\it a Lyapunov
stable point}. If all solutions that start out near the  equilibrium
point $y_f$ converge to $y_f$, then the fixed point $y_f$ is {\it an
asymptotically stable} one. Asymptotic stability of fixed point  means
that solutions that start close enough to the equilibrium not only
remain close enough but also eventually converge to the equilibrium. A
solution $y_0(t)$ of (\ref{aue}), which tends to the fixed point $y_f$,
is attractive if and only if the point $y_f$ is asymptotically stable.

The Lyapunov theorem~\cite{Lyapunov,Pontryagin} states that to prove
the stability of fixed point $y_f$ of nonlinear system (\ref{aue}) it
is sufficient to prove the stability of this fixed point for the
corresponding linearized system
\begin{equation}
\dot y=A y, \qquad A_{ik}=\frac{\partial F_i(y)}{\partial
y_k}|_{y=y_f}.
\end{equation}
The stability of the linear system means that real parts of all
solutions of the characteristic equation
\begin{equation*}
\det\left(\frac{\partial
F}{\partial y}-\lambda I\right)|_{y=y_f}=0
\end{equation*}
are negative.

In the case of {\it a hyperbolic fixed point}, i.e. the case when
 the Jacobian matrix of $F$ at the fixed point
does not have eigenvalues with zero real parts, one can use the
Hartman--Grobman theorem \cite{Grobman,Hartman,Arnold-Ilush}. This
theorem reduces the study of  the system of the first order nonlinear
equations near the hyperbolic  fixed point to the study of the behavior
of its linearization near the origin.

The case with pure imaginary eigenvalues of the Jacobian matrix of $F$
at the fixed point requires a more specific treatment
\cite{Arnold-Ilush}.

\section{Stability of isolated fixed points and kink-type solutions in one-field models with the CDM}

Let us consider the gravitational model with one scalar field $\phi$
and an arbitrary potential $V(\phi)$, described by action
(\ref{action_N}) at $N=1$. Equations (\ref{equaH2}) and (\ref{e2}) for
one-field models are as follows
\begin{equation}
\label{equsystem}
\begin{array}{l}
\displaystyle \dot H={}-3H^2+8\pi G_N(V(\phi)+\Lambda),\\
\displaystyle \dot\phi=\psi,\\
\displaystyle \dot{\psi}={}-3H\psi-\frac{1}{C}V'_{\phi}.
\end{array}
\end{equation}
This system of three first order equations is valid in the Bianchi I
metric as well as in the  FRW  one.
 Different initial values of $\sigma^2$ in (\ref{a2}) specify these different cases.

Let us define
\begin{equation}
\label{I}
  I=\frac{3}{8\pi G_N}H^2-\frac{C}{2}\psi^2-V(\phi)-\Lambda.
\end{equation}

From system (\ref{equsystem}) it is follows that the function $I$
should be a solution of the following equation:
\begin{equation}
\dot I={}-6HI. \label{equI}
\end{equation}

If the case $H(t)\equiv 0$ is excluded, then $I$ is an integral of
motion of (\ref{equsystem}) if and only if $I=0$. From (\ref{a2}) we
see that
\begin{equation}
I=\frac{1}{16\pi G_N}\sigma^2, \label{Isigma}
\end{equation}
so, $I$ is an integral of motion only at $\sigma^2=0$, i.e. in the
FRW metric. From (\ref{equI}) and (\ref{Isigma}) it follows that
equation (\ref{equvartheta}) is a consequence of (\ref{equsystem}).

We are interested in the stability of kink and lump solutions, namely,
we consider such solutions in which the Hubble parameter tends to a
finite value at $t\rightarrow+\infty$. In this case $\phi(t)$ tends to
a finite value as well. Thus, there exists a fixed point
$y_f\equiv(H_f^{\vphantom{27}},\phi_f^{\vphantom{27}},\psi_f^{\vphantom{27}})$,
which corresponds to $t=+\infty$. We consider the stability of
isotropic solutions only, so $\sigma^2_f=0$ and $\dot\beta_{i_f}=0$. It
is easy to see that
\begin{equation}
\psi_f=0, \qquad V'_{\phi}(\phi_f^{\vphantom{27}})=0, \qquad
H_f^2=\frac{8}{3}\pi G_N\left(\Lambda+V(\phi_f^{\vphantom{27}})\right).
\end{equation}

To analyse the stability of $y_f$, we present solutions as follows:
\begin{subequations}
\label{epsser}
\begin{equation}
 H=H_f + \varepsilon h(t) +{\cal
O}(\varepsilon^2)
\end{equation}
\begin{equation}
\phi= \phi_f + \varepsilon \varphi(t) +{\cal O}(\varepsilon^2)
\end{equation}
\begin{equation}
 \psi=
 \varepsilon \chi(t)+{\cal
O}(\varepsilon^2),
\end{equation}
\begin{equation}
 \dot\beta_i= \varepsilon \zeta_i(t)+{\cal
O}(\varepsilon^2),
\end{equation}
\end{subequations}
where $\varepsilon$ is a small parameter. To first order in
$\varepsilon$ we obtain the following system of equations:
\begin{subequations}
\label{linsys}
 \begin{equation}
\label{h-eq} \dot h(t)={}-6H_fh(t),
\end{equation}
\begin{equation}
\label{linsys-1}\dot\varphi(t)=\chi(t),
\end{equation}
\begin{equation}
\label{linsys-2}
\dot\chi(t)={}-3H_f\chi(t)-\frac{1}{C}V''_{\phi}(\phi_f)\varphi.
\end{equation}
\end{subequations}
Equation (\ref{h-eq}) has the solution
\begin{equation}
 h(t)=b_0e^{-6H_ft},
 \label{ht}
 \end{equation}
where $b_0$ is a constant.

From (\ref{linsys-1})--(\ref{linsys-2}) we obtain the following
solutions:
\begin{itemize}
\item at $\displaystyle V''_\phi\left(\phi_f^{\vphantom{27}}\right)\neq
0$ and $\displaystyle
V''_\phi\left(\phi_f^{\vphantom{27}}\right)\neq\frac{9C}{4}H_f^2$,
\begin{equation}
\label{linsol} \varphi(t)=D_1e^{{}-3\left(H_f^{\vphantom{27}}+
\sqrt{H_f^2-\frac{4}{9C}V''_\phi(\phi_f^{\vphantom{27}})}\right)t/2}
+D_2e^{{}-3\left(H_f^{\vphantom{27}}-
\sqrt{H_f^2-\frac{4}{9C}V''_\phi(\phi_f^{\vphantom{27}})}\right)t/2},
\end{equation}
\item at $\displaystyle
V''_\phi\left(\phi_f^{\vphantom{27}}\right)=\frac{9C}{4}H_f^2$,
\begin{equation}
\varphi(t)=e^{-3H_f^{\vphantom{27}} t/2}(D_1+D_2t),
\end{equation}
\item at $\displaystyle V''_\phi\left(\phi_f^{\vphantom{27}}\right)=0$,
\begin{equation}
\varphi(t)=\tilde{D}_1-\frac{1}{3H_f^{\vphantom{27}}}D_2e^{-3H_f^{\vphantom{27}}
t},
\end{equation}
\end{itemize}
where $\tilde{D}_1$, $D_1$, and $D_2$ are arbitrary constants.

Using the Lyapunov theorem we state that fixed point $y_f$ is
asymptotically stable and, therefore, the exact kink-type or lump-type
solution $y_0(t)$ is stable if
\begin{equation}
\frac{V''_\varphi\left(\phi_f^{\vphantom{27}}\right)}{C}>0\quad
\mbox{and}\quad H_f^{\vphantom{27}}>0. \label{Stabcond}
\end{equation}

Namely, $y_f$ is
\begin{itemize}
\item a stable focus at
$\frac{V''_\phi\left(\phi_f^{\vphantom{27}}\right)}{C}>\frac{9}{4}H_f^2$,

\item a stable node at
$\frac{9}{4}H_f^2>\frac{V''_\phi\left(\phi_f^{\vphantom{27}}\right)}{C}>0$,

\item  a stable improper node at
$V''_\phi\left(\phi_f^{\vphantom{27}}\right)=\frac{9C}{4}H_f^2$.

\end{itemize}

From (\ref{equbeta}) we obtain

\begin{equation}
\zeta_i(t)=C_ie^{-3H_ft},
\end{equation}
where $C_i$ are constants. If $H_f>0$, then $\zeta_i$ and $\sigma^2$
tend to zero at $t\rightarrow\infty$. Thus the obtained
 conditions (\ref{Stabcond}) are sufficient to prove the
stability of isotropic fixed points both in the Bianchi I and in FRW
metrics.

At $V''_\phi\left(\phi_f \right)=0$ or  $H_f=0$ we need an
additional analysis of stability, because the Lyapunov theorem does
not state the correspondence of the behavior of solutions to the
initial system (\ref{equsystem}) and the obtained linear system
(\ref{linsys}).

At $H_f<0$ the fixed point $y_f$ is unstable. Note that both $h(t)$ and
$\zeta_i(t)$, as well as $\varphi(t)$, tend to infinity at $H_f<0$.

Let us introduce the CDM into our model with a scalar field. Adding to
system (\ref{equsystem}) the CDM energy density $\rho_m$ and the
corresponding equation (\ref{rhoequ2}), we get
\begin{equation}
\begin{array}{l}
\displaystyle \dot H={}-3H^2+4\pi G_N\left(2V(\phi)+2\Lambda+\rho_m\right),\\
\displaystyle \dot\phi=\psi,\\
\displaystyle \dot{\psi}={}-3H\psi-\frac{1}{C}V'_{\phi},\\
\displaystyle\dot\rho_m={}-3H\rho_m.
\end{array}
 \label{eomlocalcdm}
\end{equation}
  Let us
consider the possible fixed points of system (\ref{eomlocalcdm}). From
the last equation of this system, it follows that  at the fixed point
we have either $H_f=0$ or $\rho_{mf}=0$. Substituting  (\ref{epsser})
and
\begin{equation}
\rho_{m}(t)=\rho_{mf}+\varepsilon\tilde{\rho}_{m}(t)+{\cal
O}(\varepsilon^2),
\end{equation}
into the system (\ref{eomlocalcdm}), we obtain the following system in
first order to $\varepsilon$:
\begin{subequations}
\label{firstorderCDM}
\begin{equation}
\dot{\tilde{\rho}}_m(t)={}-3H_f\tilde{\rho}_m(t)-3\rho_{mf}h(t),
\end{equation}
 \begin{equation}
 \dot h(t)={}-6H_fh(t)+8\pi G_N\tilde{\rho}_m(t),
\end{equation}
\begin{equation}
\dot\varphi(t)=\chi(t), \qquad \qquad \qquad \qquad { \ \ }
\end{equation}
\begin{equation}
\dot\chi(t)={}-3H_f\chi(t)-\frac{1}{C}V''_{\phi}(\phi_f)\varphi.
\end{equation}
\end{subequations}

It is easy to see that the third and fourth equations of
(\ref{firstorderCDM}) coincide with the corresponding equations of
system (\ref{equsystem}). Therefore, the case $H_f=0$ can not be
analysed by the Lyapunov theorem. Let us prove that condition
(\ref{Stabcond}) is sufficient for the stability of fixed points for models
with the CDM. First of all, from $H_f\neq 0$ it follows that
$\rho_{mf}=0$. Solving the first and second equations of
(\ref{firstorderCDM}), we obtain
\begin{equation}
\tilde{\rho}_m(t)=b_1e^{-3H_ft},\qquad
h(t)=b_0e^{-6H_ft}+\frac{b_1}{3H_f}e^{-3H_ft},
\end{equation}
where $b_1$ is an arbitrary constant.

We come to the conclusion that if conditions (\ref{Stabcond}) are satisfied, then the solution, which
is stable in the model without the CDM, is stable with respect to the CDM energy density
fluctuations as well.

\section{Connections between the first order corrections to isotropic
solutions in the FRW and Bianchi I metrics}

In the previous section we studied one-field models and the first
corrections near a fixed point. In this section we consider the first
corrections of an arbitrary isotropic solution.

We consider an $N$-field cosmological model, which is described by
action (\ref{action_N}) and the Einstein equations
(\ref{b})--(\ref{e}). In this section we do not assume that the
isotropic solution tends to a fixed point. We do not prove the
stability of solutions, we only analyse the first corrections in the
FRW and Bianchi I metrics. To apply the Lyapunov theorem it was
convenient to consider functions $H$ and $\dot\beta_i$ instead of
$H_i$. In this section we return to functions $H_i$.

To study the stability of this solution, we present solutions whose
initial conditions are close to the isotropic one, in the following
form: \bea \label{j} H_i(t)&=&H_0(t)+\varepsilon h_i(t)+{\cal
O}(\varepsilon^2),\\
\label{m} \phi_k(t)&=&\phi_{0k}(t)+\varepsilon
\varphi _k(t)+{\cal
O}(\varepsilon^2),\\
\psi_k(t)&=&\psi_{0k}(t)+\varepsilon \chi_k(t)+{\cal
O}(\varepsilon^2),\\
\label{rho_m} \rho_m(t)&=&\rho_{m0}(t)+\varepsilon
\tilde{\rho}_m(t)+{\cal O}(\varepsilon^2),
 \eea
 where $i=1,2,3$ and $k=1,\dots,N$.
From (\ref{a})--(\ref{e})  we obtain to zero order in $\varepsilon$ the
system of  Einstein equations and equations of motion in the FRW
metric. To first order in $\varepsilon$ we have the following system:
\bea
\label{a1} \dot \varphi_{k}&\!=\!&\chi_{k},\\
 \dot \chi_{k}&\!=\!&{}-(h_1+h_2+h_3)\psi_{0k}
-3H_0\chi_{k}-\frac{1}{C_k}\sum_{m=1}^N
V''_{\phi_k\phi_m}\left(\phi_0\right)\varphi_{m},\\
\dot{\tilde{\rho}}_m&\!=\!&{}-(h_1+h_2+h_3)\rho_{m0}-3H_0\tilde{\rho}_m,\\
\label{b1} \dot h_1+\dot h_2&\!=\!&{}-3H_0(h_1+h_2)+8\pi
G_N\sum_{k=1}^N\left(V'_{\phi_k}\left(\phi_{0}\right)\varphi_{k} -
C_k\dot {\phi}_{0k}\chi_{k}\right),
\\
\label{k1} \dot h_1+\dot h_3&\!=\!&{}-3H_0(h_1+h_3)+8\pi
G_N\sum_{k=1}^N\left(V'_{\phi_k}\left(\phi_{0}\right)\varphi_{k} -
C_k\dot {\phi}_{0k}\chi_{k}\right),
\\
\label{d1}\dot h_2+\dot h_3&\!=\!&{}-3H_0(h_2+h_3)+8\pi
G_N\sum_{k=1}^N\left(V'_{\phi_k}\left(\phi_{0}\right)\varphi_{k} -
C_k\dot {\phi}_{0k}\chi_{k}\right). \eea

From equations (\ref{b1})--(\ref{d1}) we get \bea
\label{equh12} \dot h_1(t)-\dot h_2(t)+3H_0(t)(h_1(t)-h_2(t))\!\!&\!=\!&\!\!0,\\
\label{equh13} \dot h_1(t)-\dot
h_3(t)+3H_0(t)(h_1(t)-h_3(t))\!\!&\!=\!&\!\!0,
 \eea
and we also have
\begin{equation}
 H_0(h_1+h_2+h_3)=4\pi G_N\sum_{k=1}^N\left(C_k\dot
\phi_{0k}\dot \varphi_{k}+
V'_{\phi_k}\left(\phi_{0}\right)\varphi_{k}\right).
\end{equation}

\begin{theor}
{ \ }

Let $H_0(t)$ be a smooth function bounded at all finite values of time
and $\int \limits_{0}^\infty H_0(\tau)d\tau$ be bounded from below, in
other words, this integral is equal to either a finite number or plus
infinity. Functions $h_1(t)$, $h_2(t)$, $h_3(t)$, $\tilde{\rho}_m(t)$,
and $\varphi_{k}(t)$, which are solutions of (\ref{a1})--(\ref{d1}),
are bounded if and only if isotropic solutions, namely, solutions,
which satisfy the condition $h_1(t)=h_2(t)=h_3(t)$, are bounded.
\end{theor}

{\bf Proof.} It is trivial that if the full set of solutions includes
only boundary functions, then any subset which satisfies an additional
condition includes only boundary functions. Let us prove that the
boundedness of isotropic solutions is not only a necessary condition,
but also a sufficient one.

From equations (\ref{equh12}) and (\ref{equh13}) we obtain:
\begin{equation}
\label{h12} h_1(t)-h_2(t)=(h_1(0)-h_2(0))e^{-3\int
\limits_{0}^tH_0(\tau)d\tau},\quad
h_1(t)-h_3(t)=(h_1(0)-h_3(0))e^{-3\int \limits_{0}^tH_0(\tau)d\tau}.
\end{equation}

So we obtain that if the integral $\int \limits_{0}^tH_0(\tau)d\tau$ is
uniformly bounded from below, then anisotropy is bounded at all $t$.
Note that in the most of cosmological models $H_0(t)>0$ for all $t>0$
and the anisotropy tends to zero at $t\rightarrow\infty$.

Using (\ref{h12}), one can express $h_2(t)$ and $h_3(t)$ via $h_1(t)$
and reduce system (\ref{b1})--(\ref{d1}) to one equation. System
(\ref{a1})--(\ref{d1}) takes the following form:
\bea \label{aa1}
2H_0\left(3h_1-C_0e^{-3\int \limits_{0}^tH_0(\tau)d\tau}\right)&=&8\pi
G_N\left(\sum_{k=1}^N C_k\dot \phi_{0k}\dot \varphi_{k}+\sum_{k=1}^N
V'_{\phi_k}\left(\phi_{0}\right)\varphi_{k}\right),
\\
\label{bb1} 2\dot h_1+6H_0h_1&=&8\pi G_N\left(\sum_{k=1}^N
V'_{\phi_k}\left(\phi_{0}\right)\varphi_{k}-\sum_{k=1}^N C_k\dot
\phi_{0k}\dot \varphi_{k}\right), \eea where
$C_0=2h_1(0)-h_2(0)-h_3(0)$.

Let us introduce a new function,
\begin{equation}
\label{h0}
    h_0(t)\equiv h_1(t)-\frac{C_0}{3}e^{-3\int
\limits_{0}^tH_0(\tau)d\tau}.
\end{equation}
It is easy to check that
\begin{equation}
\label{3h} 3h_0(t)= h_1(t)+ h_2(t)+ h_3(t).
\end{equation}

System (\ref{aa1})--(\ref{bb1}) in terms of $h_0$ and $\varphi_{k}$
coincides with the system of equations (\ref{a1})--(\ref{d1}) with
$h_1(t)=h_2(t)=h_3(t)=h_0(t)$. In other words, we obtain that the
functions $\varphi_{k}(t)$ in the Bianchi I and FRW metrics are the
same. Functions $h_1(t)$, $h_2(t)$, and $h_3(t)$ differ from the
correction for the Hubble parameter $h_0(t)$ on a finite value. Thus
the theorem is proven.

Note that Theorem 1 connects the stability properties of the FRW and
Bianchi I metrics not only for solutions which tend to a fixed point,
but also for solutions which tend to infinity at $t\rightarrow\infty$.
Examples of such solutions in the cosmological models are presented in
Sections $5$ and $6$.

\section{Stability  of  solutions  in the  $k$-essence  model  in  the  Bianchi I  metric}
\subsection{First order corrections}
Let us consider the first order corrections in the $k$-essence model.
Substituting
\begin{equation}
{\cal E}={\cal E}_0+\varepsilon{\cal E}_1+{\cal O}(\varepsilon^2),
\qquad {\cal P}={\cal P}_0+\varepsilon {\cal P}_1+{\cal
O}(\varepsilon^2),
\end{equation}
in (\ref{equKes1})--(\ref{equKes4}) and expanding (\ref{j}) to first
order in $\varepsilon$, we obtain the following system:
\begin{eqnarray}
\label{ke1} 2H_0(\dot h_1+\dot h_2+\dot h_3)&\!=\!&{}8\pi G_N{\cal
E}_1,\\
\label{ke2} \dot h_1+\dot h_2&\!=\!&{}-3H_0(h_1+h_2)-8\pi
G_N{\cal P}_1,
\\
\label{ke3} \dot h_1+\dot h_3&\!=\!&{}-3H_0(h_1+h_3)-8\pi G_N{\cal
P}_1,
\\
\label{ke4}\dot h_2+\dot h_3&\!=\!&{}-3H_0(h_2+h_3)-8\pi G_N{\cal P}_1.
\end{eqnarray}
It is easy to see that equations (\ref{equh12}) and (\ref{equh13}) can
be obtained from (\ref{ke2})--(\ref{ke4}), therefore, formula
(\ref{h12}) is valid for solutions of system (\ref{ke1})--(\ref{ke4}).
So, it is useful to introduce $h_0$ by formula (\ref{h0}). Because of
(\ref{3h}) we obtain that system (\ref{ke1})--(\ref{ke4}) in terms of
$h_0$, ${\cal E}_1$, and ${\cal P}_1$ coincides with the corresponding
equations in the FRW metric, so if $H_0$ satisfies the conditions of
Theorem 1, then solutions of (\ref{ke1})--(\ref{ke4}) are bounded if
and only if isotropic solutions, namely, solutions which satisfy the
condition $h_1(t)=h_2(t)=h_3(t)$, are bounded. This means that it is
sufficient to calculate the first order corrections for the given
background solutions in the FRW metric to describe their behavior in
the Bianchi I metric.

\subsection{Example}

Let us consider the following example:
\begin{equation}
\varrho_q(\Phi)=C_1+B\Phi^2,\qquad p_q(\Phi)=C_2-B\Phi^2,\qquad
M(\Phi)=M_0,
\end{equation}
where $B$, $C_1$, $C_2$, and $M_0$ are constants. The Friedmann
equations are
  \bea
 3H^2&=&8\pi G_N({\cal E}+\Lambda),\\
\dot H&=&{}-4\pi G_N({\cal E}+{\cal P}). \label{equH2} \eea One can
check that the following  exact solution exists:
\begin{equation}
\Phi_0(t)=t,\qquad H_0(t)={}-4\pi G_N(C_1+C_2)t, \label{kesol}
\end{equation} if
\begin{equation}
    C_1={}-\Lambda,\qquad B=6\pi G_N(C_1+C_2)^2.
\end{equation}
Let us analyse the stability of the exact solution,
\begin{equation}
    \Phi=\Phi_0(t)+\varepsilon\Psi(t),\qquad
    H=H_0(t)+\varepsilon h(t).
\end{equation}
The equations for the first order fluctuations,
\bea \dot
h(t)&=&{}-8\pi G_N(C_1+C_2+2M^4)\dot\Psi(t),\\
\dot\Psi(t)&=&{}-12\pi
G_N\frac{t(C_1+C_2)}{(C_2+C_1+4M^4)}\left[(C_1+C_2)\Psi(t)+2m_p^2h(t)\right],
 \eea
have the following general solution:
\begin{equation}
    h(t)=d_1e^{6\pi G_N(C_1+C_2)t^2}+d_2,
\end{equation}
\begin{equation}
    \Psi(t)={}-\frac{1}{8\pi G_N}\left(\frac{d_1}{C_2+C_1+2M_0^4}e^{6\pi G_N(C_1+C_2)t^2}
    +\frac{2d_2}{C_1+C_2}\right),
\end{equation}
where $d_1$ and $d_2$ are arbitrary numbers.

If $C_1+C_2<0$, then $H_0>0$ at $t>0$ and the exact solution is stable
in the sense that the first corrections are bounded functions. Similar
solutions, obtained from the SFT inspired model, are considered in
Subsection 6.5.

\section{Examples of isotropic stable soluti\-ons in the SFT inspired mo\-dels}

\subsection{String field theory inspired cosmological models}

 An interest in cosmological
models coming from  open string field
theories~\cite{IA} is caused by a possibility to get
solutions rolling from a perturbative vacuum to the true one.
 When all other massive fields are
integrated out by means of equations of motion, the open string tachyon
acquires a nontrivial potential with a nonperturbative minimum. For the
open fermionic NSR string with the GSO$-$ sector~\cite{NPB} in a
reasonable approximation, one gets the Mexican hat potential for the
tachyon field (see~\cite{SFT-review} for a review).  Rolling of the
tachyon from the unstable perturbative extremum towards this minimum
describes, according to the Sen conjecture~\cite{SFT-review}, the
transition of an unstable D-brane to a true vacuum.  In fact one gets a
nonlocal potential with a string scale as a parameter of nonlocality.
After a suitable field redefinition the potential becomes local,
meanwhile, the kinetic term becomes nonlocal. This nonstandard kinetic
term  has a so-called phantomlike  behavior and can be approximated by
a phantom kinetic term. Rolling solutions are particular examples of
kink-type solutions.

It is also interesting to study lump-type solutions, which in particular
have no time singularity. In was advocated in \cite{AJ, Calcagni} that
such solutions are also available in the SFT inspired models.

In this section we consider the stability of the kink-type and
lump-type solutions for the SFT inspired cosmological
models~\cite{AKV,AJ,AJV2} under perturbations in the Bianchi I metric.

In~\cite{AKV,AJ} we have considered the SFT inspired phantom models
with high degree polynomial potentials. We consider the stability of
the obtained exact solutions in the next two subsections.

In~\cite{AJV2} we have considered a nonlocal cosmological model with
quadratic potential and obtained that exact solutions of this model are
solutions of local models with quadratic or zero potential. In
Subsections 6.4 and 6.5 we analyse the stability of these solutions in
massless and massive cases correspondingly. The exact solutions in the
massive case are similar to solutions (\ref{kesol}) in the $k$-essence
model.

In the examples we use a dimensionless parameter  $m_p^2 \sim
M_p^2=1/(8\pi G_N)$. The coefficient of proportionality arises when we
construct  effective cosmological models from the original SFT action
(see~\cite{AKV,AJ,AJV2} for details). For convenience, we write the
Einstein equations for the SFT inspired cosmological models in the
following form:
\begin{equation}
\begin{array}{rcl}
\displaystyle \dot H&\displaystyle=&\displaystyle{}-\frac32
H^2-\frac{1}{2m_p^2} \left(\frac{C
\psi^2}{2}-V(\phi)-\Lambda\right),\\[2.7mm]
\displaystyle \dot\phi&\displaystyle=&\displaystyle\psi,
\\[2.7mm]
\displaystyle \dot\psi&\displaystyle=&\displaystyle
{}-3H\psi-\frac{1}{C}V'_{\phi}(\phi).
\end{array}
 \label{eomSFTinsp}
\end{equation}
We also have
\begin{equation}
  3m_p^2H^2-\frac{C}{2}\phi^2-V(\phi)=\Lambda.
\end{equation}

\subsection{Model with a kink solution and the sixth degree potential}

An exact solution to the Friedmann equations with a string inspired
phantom scalar matter field has been constructed in~\cite{AKV} (see
also~\cite{AKVCDM}). The notable features of the model are a phantom
sign of the kinetic term ($C=-1$) and a special polynomial form of the
effective tachyon potential:
\begin{equation}
V(\phi)= \frac12 \left(1-\phi^2\right)^2 +\frac{1}{12m_p^2}\,
\phi^2\left(3-\phi^2\right)^2. \label{Vphi}
\end{equation}

Note that this potential has been used in the string gas
cosmology~\cite{McInnes}.

System (\ref{eomSFTinsp}) has the following exact kink-type
solution~\cite{AKV}:
\begin{equation}
\label{ES} \phi_0(t)=\tanh(t),\qquad
H_0(t)=\frac{1}{2m_p^2}\tanh(t)\left(1-\frac{1}{3}\tanh(t)^2\right).
\end{equation}

Let us analyse the stability of this solution. At $t\rightarrow\infty$
solution (\ref{ES}) tends to a fixed point,
\begin{equation}\label{FP1}
H_f=\frac{1}{3m_p^2}, \qquad \phi_f=1.
\end{equation}
It is easy to see that
\begin{equation}
V'_\phi(1)=0,\qquad V''_{\phi\phi}(1)=2\left(2-\frac{1}{m_p^2}\right).
\end{equation}
Using (\ref{Stabcond}), we obtain that solution (\ref{ES}) is
attractive in the Bianchi I metric at $m_p^2 < 1/2$. Note that this
solution is stable with respect to small fluctuations of the initial
value of the CDM energy density as well.

In~\cite{AKV} we have showed that the first corrections $\varphi(t)$
and $h(t)$ satisfy the following system:
\begin{equation}
\begin{array}{@{}rcl@{}}
\displaystyle\dot h&\displaystyle=&\displaystyle\frac{1}{m_p^2}\Bigl(1-
\tanh(t)^2\Bigr)\dot{\varphi},\\[3.7mm]
\displaystyle\dot{\varphi}&\displaystyle=&
\displaystyle\frac{\left(3-4m_p^2
+4(m_p^2-1)\tanh(t)^2+\tanh(t)^4\right)\tanh(t)}
{2m_p^2\left(1-\tanh(t)^2\right)}
\varphi-{}\\[2.7mm]&\displaystyle-&\displaystyle
 \frac{\left(3-\tanh(t)^2\right)\tanh(t)}{1- \tanh(t)^2}h,\\
\end{array}
\label{equeps}
\end{equation}
and have the following explicit form:
\begin{equation}
\begin{split}
\varphi(t)&\displaystyle=2m_p^2C_1\left(1-\tanh(t)^2\right)+{}\\
&{}+2m_p^2 C_2\frac{2 J(t)+(\cosh(2t)-1)(\cosh(t))^{2-\frac{1}{m_p^2}}
e^{\left(\frac{1}{2m_p^2(\cosh(2t)+1)}\right)}}{\cosh(2t)+1},\\
h(t)&=C_1\left(1-\tanh(t)^2\right)^2-\frac{4m_p^2{C}_2 J(t)}
{(\cosh(2t)+1)^2},\\
\end{split}
\label{phi1H1}
\end{equation}
where $C_1$ and $C_2$ are arbitrary constants,
\begin{equation*}
 J(t)=\int_0^t \!\sinh\left(\tau\right) \left(
\cosh(\tau)\right)^{1-1/m_p^2}\left(
2\left(2m_p^2-1\right)\cosh(\tau)^2-1 \right)
e^{\frac{1}{4m_p^2\cosh(\tau)^2}}d\tau.
\end{equation*}

It is easy to see that if $m_p^2>1/2$ then at $C_2\neq 0$ the function
$\varphi(t)$ tends to infinity as $t\rightarrow\infty$ and, therefore,
solution (\ref{ES}) is not stable. At $m_p^2=1/2$ we obtain from
(\ref{phi1H1}) that
\begin{equation*}
\begin{split}
h(t)&=\left(\tanh(t)^2-1\right)^2\left(C_1-C_2J_2\right),\\
\varphi(t)&=-\left(\tanh(t)^2-1\right)\left(C_1-C_2J_2\right)
-\frac{1}{2}C_2e^{-\tanh(t)^2/2},
\end{split}
\end{equation*}
where $J_2=\int_0^t \!{e^{-\tanh(\tau)^2/2}}\tanh(\tau){d\tau}$. Thus,
$\varphi(t)$ and $h(t)$ are bounded functions at $m_p^2= 1/2$.

The functions  $h_i$ have the form
\begin{equation}
    h_i(t)=h(t)+\tilde{C_i}\,e^{{}-\frac{\tanh^2(t)}{4m_p^2}}
    \left(1-\tanh^2(t)\right)^{1/\left(2m_p^2\right)},
\end{equation}
where $\tilde{C_i}$ are real constants, $i=1,2,3$, which satisfy the
following relation:
\begin{equation}
\label{3C0}
\tilde{C_1}+\tilde{C_2}+\tilde{C_3}=0.
\end{equation}

We conclude that exact solutions obtained in~\cite{AKV} are stable in
the Bianchi~I metric at $m_p^2< 1/2$ and unstable at  $m_p^2> 1/2$. The
case of $m_p^2= 1/2$ needs a more detailed analysis. The first
corrections are bounded.

\subsection{Model with a lump solution}

In the previous subsection kink solutions were considered. In this
subsection we consider the stability of a lump solution in the
model~\cite{AJ} which is motivated by a description of D-brane decay
within the string field theory framework. We take the one-field
cosmological model with the potential
\begin{equation}
V(\phi)=2(1-\phi)\phi^2-\frac{4(\phi-1)^3(2+3\phi)^2}{75 m_p^2}
\end{equation}
and $C=-1$. The Friedmann equations (\ref{eomSFTinsp}) have the
following exact solution~\cite{AJ}:
\begin{subequations}
\begin{equation}
\phi_0 = \sech^2(t),
\end{equation}
\begin{equation}
\label{6H0}
H_0 =\frac{2(3+2\cosh(t)^2)\tanh^3(t)}{15 m_p^2\cosh(t)^2}.
\end{equation}
\label{LS}
\end{subequations}
At $t\rightarrow\infty$ solution (\ref{LS}) tends to a fixed point:
\begin{equation}\label{FP2}
H_f=\frac{4}{15m_p^2}, \qquad \phi_f=0.
\end{equation}
It is easy to see that
\begin{equation}
V'_\phi(0)=0,\qquad V''_{\phi\phi}(0)=4\left(1-\frac{2}{5m_p^2}\right).
\end{equation}
Using (\ref{Stabcond}), we obtain that solution (\ref{LS}) is
attractive in the Bianchi I metric at $m_p^2 < 2/5$. In~\cite{AJ} the
authors consider a model without the CDM, at the same time, the results
of Section 2 show that solution (\ref{LS}) is stable with respect to
small fluctuations of the initial value of the CDM energy density as
well.

Let us perturb the Friedmann equations in the standard way,
\begin{equation}
H=H_0(t)+\epsilon h(t), \quad \phi=\phi_0(t)+\epsilon \varphi(t).
\end{equation}

To first order in $\epsilon$ we have the following system of equations:
\begin{equation}
\label{ljequ}
\begin{array}{l}
\displaystyle\dot h+\frac{2}{m_p^2} \sech^2(t) \tanh ^2(t)
\dot \varphi=0,\\
\displaystyle\frac{1}{m_p^2}\left(\frac{4}{5}(4+\cosh(2 t))\sech^2(t)\tanh^3(t) h+ (6
\sech^4(t) -4\sech^2(t))\varphi\right)-{}\\
\displaystyle{}-\frac{4(2+3\sech^2(t))^2 }{25
m_p^2}(\tanh^4(t)-2\tanh^6(t))\varphi +2 \sech^2(t)
\tanh(t) \dot\varphi=0.
\end{array}
\end{equation}

System (\ref{ljequ}) has the following solutions:
\begin{equation*}
\varphi=\frac{1}{2\sinh(t) \cosh^3 (t)} \left( 5 C_2 m_p^2
\cosh(t)^{(\frac{-4+30 m_p^2}{5 m_p^2})}
 e^{\left(\frac{2 \cosh^2 (t)-3}{10m_p^2 \cosh^4 (t)}\right)}
 - 2C_1 \cosh^2 (t)-{}\right.
\end{equation*}
\begin{equation*}
{} - 2 C_2 \int  \frac{1}{\sinh^3 (t)}(-15m_p^2 \cosh^4 (t) +10 m_p^2
\cosh^6 (t) +8\cosh^2(t)-6-4 \cosh^6 (t)+2\cosh^4 (t))\times
\end{equation*}
\begin{equation*}
\left.\times\cosh (t)^{(\frac{-4+5 m_p^2}{5 m_p^2})}
 e^{\left(\frac{2 \cosh^2 (t)-3}{10m_p^2 \cosh^4(t)}\right)}(\cosh^2(t)-m_p^2) dt +2 C_1
 \right),
\end{equation*}

\begin{equation*}
h= \frac{16(\cosh (2t)-1)}{\cosh (6 t) + 6 \mathbf{\cosh} (4t) + 15
\cosh (2t) +10}\Bigl( C_1+C_2 \int \frac{\cosh (t)^{(\frac{-4+5
m_p^2}{5 m_p^2})}
 e^{\left(\frac{2 \cosh^2 (t)-3}{10m_p^2 \cosh^4 (t)}\right)}}{\sinh^3
 (t)}\times
 \end{equation*}
\begin{equation*}
\times[ -15 m_p^2 \cosh^4 (t) + 10 m_p^2 \cosh^6 (t)+ 8 \cosh^2 (t)-6 -
4 \cosh^6 (t) +2 \cosh^4 (t) ]dt\Bigr).
\end{equation*}
Using (\ref{6H0}) we get
\begin{equation}
 h_i=h+\tilde{C}_i
\cosh(t)^{{}-\frac{4}{5m_p^2}}e^{\frac{\sinh(t)^2\left(3+\cosh(t)^2\right)}{10m_p^2\cosh(t)^4}},
\end{equation}
 where $\tilde{C}_i$ are arbitrary real constants which satisfy
(\ref{3C0}).

It is easy to verify that $h(t)$ and $h_i(t)$ are bounded functions for
any values of the parameters. Taking into account that
\begin{equation}
\lim_{t->\infty} \exp\left(\frac{2\cosh^2(t)-3}{10m_p^2
\cosh^4(t)}\right)=1,
\end{equation}
we obtain that $\varphi$ is bounded at $m_p^2 \leqslant 2/5$ and
unbounded at $m_p^2 > 2/5$. The stability in the case of $m_p^2=2/5$
cannot be analysed without using high order corrections.

In the examples considered in this and the previous subsections, the
potentials depend on $m_p^2$, which results in the fact that extrema of
potentials are either minima, maxima, or inflection points. In the next
two examples we consider the opposite case, when potentials do not
depend on $m_p^2$.

\subsection{Model with    a    massless phantom field}

At present nonlocal cosmological models are being studied very
actively~\cite{IA}, \cite{AJ}--\cite{KV}. In this and the next
subsections we consider the stability of solutions of local models
which correspond to the nonlocal model with a quadratic potential.
These solutions have been presented in~\cite{AJV2}, where the method of
localizing the nonlocal model with a quadratic potential has been
proposed.

Let us consider the one-field model with zero potential, $V(\phi)=0$.
From the Friedmann equations (\ref{eomSFTinsp}) we obtain
\bea 3H^2&=&
\frac{C}{2m_p^2}\dot\phi^2+\frac{\Lambda}{m_p^2},\\
\dot H&=&{}-\frac{C}{2m_p^2}\dot\phi^2. \label{eomprholocal1} \eea

At $\Lambda>0$ and $C<0$ there are the following real solutions:
\begin{equation}
\begin{array}{l}
\displaystyle \phi_0(t)={}\pm\sqrt{{}-\frac{2m_p^2}{3C}}\arctan
\left(\sinh\left(\sqrt{\frac{3\Lambda}{m_p^2}}(t-t_0)\right)\right)+C_1,\\
\displaystyle
H_0(t)=\sqrt{\frac{\Lambda}{3m_p^2}}\tanh\left(\sqrt{\frac{3\Lambda}
{m_p^2}}(t-t_0)\right),
\end{array}
\label{phi0H0V0}
\end{equation}
where $t_0$ and $C_1$ are arbitrary real constants.

Let us consider the stability of the solution ($H_0,\phi_0$).
Substituting $H_0$ and $\phi_0$ into (\ref{j})--(\ref{m}), to first
order in $\varepsilon$ we obtain
\begin{equation}
\begin{array}{l} \displaystyle
\varphi(t)={}\pm\frac{2m_p^2\sqrt{2}e^{2\sqrt{3m_p^2\Lambda}(t-t_0)/m_p^2}}
{\sqrt{-C\Lambda}\left(
e^{2\sqrt{3m_p^2\Lambda}(t-t_0)/m_p^2}+1\right)}C_3 +C_2,\\[12.72mm]
\displaystyle
h(t)=\frac{2C_3}{\cosh\left(2\frac{\sqrt{3m_p^2\Lambda}}{m_p^2}(t-t_0)\right)+1},
\end{array}
\end{equation}
where $C_2$ and $C_3$ are arbitrary real constants. It is obvious, that
functions  $h(t)$ and $\varphi(t)$ are bounded. In the Bianchi I metric
we have
\begin{equation}
    h_i(t)=h(t)+\tilde{C}_i\sqrt{1-\tanh^2
    \left(\frac{\sqrt{3m_p^2\Lambda}}{m_p^2}(t-t_0)\right)},
\end{equation}
where real constants $\tilde{C}_i$, $i=1,2,3$, satisfy the following
relation:
\begin{equation}
\tilde{C}_1+\tilde{C}_2+\tilde{C}_3=0.
\end{equation}

Thus, we have obtained that the kink-type solutions (\ref{phi0H0V0}) in
the Bianchi I metric have the bounded first corrections.

\subsection{Model with  a quadratic potential
and the cosmological constant}

Let us consider the model of a scalar field with a quadratic potential
and the cosmological constant. In this case the Friedmann equations are
\begin{equation}
\label{by} H^2=\frac{8\pi
G_N}{3}\left(\frac{C}{2}\dot{\phi}^2+\frac{B}{2}\phi^2+\Lambda\right),
\end{equation}
\begin{equation}
\label{bz} \dot{H}={}-4\pi G_NC\dot{\phi}^2,
\end{equation}
where $C$ and $B$ are arbitrary nonzero real numbers.

System (\ref{by})--(\ref{bz}) has the following particular solutions:
\begin{equation}
\label{cb} H_0(t)=k_1t,\qquad
 \phi_0(t)=k_2t,
\end{equation}
where
\begin{equation}
\label{cd} k_1={}-\frac{B}{3C}, \qquad k_2^2=\frac{B}{12\pi G_N C^2}.
\end{equation}

From (\ref{cd}) it follows that the function $\phi$ is real if and only
if $B>0$. The above-mentioned solutions exist only if
\begin{equation}
\label{ce} \Lambda={}-\frac{B}{24\pi G_NC}.
\end{equation}
To analyse the stability of these exact solutions, we substitute
\begin{equation}
\label{cf} H(t)=k_1t +\varepsilon h(t)
\end{equation}
and
\begin{equation}
\label{cg} \
\phi(t)=k_2t+\varepsilon \varphi(t).
\end{equation}
 in (\ref{by}) and (\ref{bz}).

To first order in $\varepsilon$ we obtain the following system of
equations:
\begin{equation}
\label{ch} \dot\varphi(t)={}-\frac{Bt}{C}\left(\varphi(t)+\frac{1}{4\pi
G_NCk_2}h(t)\right),
\end{equation}
\begin{equation}
\label{ci} \dot{h}(t)={}-8\pi G_NCk_2\dot{\varphi}(t).
\end{equation}
Solutions of (\ref{ch})--(\ref{ci}) are
\begin{equation}
\label{ck} h(t)=\tilde{D}_1e^{\frac{B}{2C}t^2}+\tilde{D}_2,
\end{equation}
 \begin{equation}
\label{cj} \varphi(t)= {}-\frac{1}{8\pi C G_N
k_2}\left(2\tilde{D}_{1}+\tilde{D}_2e^{\frac{B}{2C}t^2}\right),
\end{equation}
where $\tilde{D}_1$ and $\tilde{D}_2$ are arbitrary constants,
\begin{equation}
k_2={}\pm\sqrt{\frac{B}{12\pi G_N C^2}}.
\end{equation}

Therefore, the functions $h(t)$ and $\varphi(t)$ are bounded at
$C/B<0$. Real solutions exist only if $B>0$, and hence, $C<0$. We come
to the conclusion that solution (\ref{cb}) can be stable (the first
corrections are bounded) only if $C<0$, in other words, $\phi(t)$ is a
phantom scalar field.  In this case $H_0(t)>0$ at $t>0$, hence $h_i$
are bounded as well. Indeed,
\begin{equation}
h_i(t)=h(t)+\tilde{C}_ie^{-3\int
\limits_{0}^tH_0(\tau)d\tau}=(\tilde{D}_{1}+\tilde{C}_i)e^{\frac{B}{2C}t^2}+\tilde{D}_{2},
\end{equation}
are bounded. Constants $\tilde{C}_i$ satisfy the relation (\ref{3C0}).

\section{Conclusion}

We have analysed the stability of isotropic solutions for the models
with NEC violation in the Bianchi I metric.

In our paper for the one-field model with the CDM we used the Lyapunov
theorem and found sufficient conditions for stability of kink-type and
lump-type solutions both in the FRW metric and in the Bianchi I metric.
The obtained results allow us to prove that the exact solutions, found
in string inspired phantom models~\cite{AKV,AJ}, are stable. A
generalization of this result to two-field models, for example, quintom
models, requires further studies and will be considered in future
investigations.

We  found the explicit form of the connection between $h_1(t)$,
$h_2(t)$, and $h_3(t)$, which define metric perturbations in the
Bianchi I metric, and $h_0$, which defines perturbations in the FRW
metric. We have proved that fluctuations for the fields and the CDM
energy density in both metric are the same. In particular, for
$H_0\geqslant 0$ the boundedness of $h_0$ is a sufficient and necessary
condition for the boundedness of  $h_1(t)$, $h_2(t)$, and $h_3(t)$.
This result is valid for both $N$-field and $k$-essence models.

Note that linear-in-time solutions for the simple models with a
quadratic potential, which have been considered in Subsection 6.5,  are
stable with respect to the first corrections only in the phantom case
(which corresponds to $H>0$). It means that NEC violation does not lead
to the instability in this sense. It gives us an intuitive reason to
expect that more complicated NEC violated models  can also have stable
isotropic solutions. This expectation has been confirmed in the models
considered in Section 6.

To conclude, our results are for the NEC violated models. If energy
conditions are satisfied, then results, similar to those obtained in
Section 4, are consequences of the Wald theorem~\cite{Wald} and its
generalizations~\cite{MossSahni,Wainwright,Kitada,Rendall04}.

Our study of the stability of isotropic solutions for the models with
NEC violation in the Bianchi I metric shows that the NEC is not a
necessary condition for classical stability of isotropic solutions.
Because of strong limits on anisotropic models from
observations~\cite{Barrow,Bernui:2005pz}, cosmological models
developing large anisotropy should be discarded. In this paper we have
shown that the models~\cite{AJ,AJV2,AKV} have stable isotropic
solutions and that large anisotropy does not appear in these models.

\section*{Acknowledgements}
The authors are grateful to Alexei A. Starobinsky for drawing their
attention to the stability problem of isotropic solutions in the
Bianchi metrics, in particular, in the Bianchi I metric. L.J. would
like to thank D. Mulryne and D. Wesley for useful discussions.

I.A., N.B. and S.V. are supported in part by a state contract from the
Russian Federal Agency for Science and Innovations No. 02.740.11.5057.
I.A. and S.V.  are supported in part by RFBR grant No. 08-01-00798 and
by the Russian Ministry of Education and Science though grant No.
NSh-795.2008.1 (I.A.) and No. NSh-1456.2008.2 (S.V.).  L.J.
acknowledges the support of the Centre for Theoretical Cosmology, in
Cambridge.

\end{document}